\documentclass{tMPH2e}

\usepackage[colorlinks = true, linkcolor = black, citecolor = black, urlcolor = blue]{hyperref}
\usepackage{breakurl}

\begin{document}
\doi{10.1080/0026897YYxxxxxxxx}
 \issn{1362–3028}
\issnp{0026–8976}
\jvol{00}
\jnum{00} \jyear{2012} 

\markboth{A.A. Louis}{Molecular Physics}

\articletype{}

\title{{\itshape Extracting short-ranged interactions from structure factors }}

\author{A.A. Louis$^{\ast}$\thanks{$^\ast$Corresponding author. Email: ard.louis@physics.ox.ac.uk}{ \\ \vspace*{6pt} \em{Rudolf Peierls Centre for Theoretical Physics, 1 Keble Road, Oxford 0X1 3NP, England}}}

\maketitle

\begin{abstract}
Inverting scattering experiments to obtain effective interparticle
interactions is generally a poorly
conditioned problem.   L. Reatto (Phil. Mag. A {\bf 58}, 37 (1986))  showed that  for atomic liquids close to the triple point,  inversions are hard because the structure closely resembles that of an equivalent hard-sphere fluid.    Here I demonstrate that at low concentrations and  for particles with short-ranged attractive potentials,  $S(k)$ also exhibits a very weak dependence on potential shape. Instead,  different potentials all generate an $S(k)$ that  closely resembles that of the Baxter model with a similar second-virial coefficient.   By contrast, in this energetic fluid regime, the inversion of an attractive interaction from real-space correlations such as  the radial distribution function $g(r)$ is well conditioned.  Nevertheless,  one may extract further  information from $S(k)$ by measuring {\em isosbestic points}, values of $k$ where the scattering
intensity $I(k)$ or the structure factor $S(k)$ is invariant to
changes in interaction-potential well-depth.  These points suggest a new
extended corresponding states principle for particles in solution
based on the packing fraction, the second osmotic virial coefficient,
and a new measure of effective potential range.
\end{abstract}

\bigskip


\section{Introduction}

Extracting effective interparticle interactions from experiments is a
central objective in chemical physics because these interactions
govern physical behaviour~\cite{Liko01}.  But like many such inverse
problems, this task is complicated. Experimental data is not perfect:
statistical fluctuations occur, and results may not be obtainable over
a complete parameter range.  Moreover, the inversion procedures are
often ill conditioned: small differences in the initial input can
result in large changes in the final output.

  This article concentrates on the determination of
(spherically symmetric) effective interparticle pair potentials
$v^{eff}(r)$ from experimental scattering data for particles in
solution.  This problem is related to better studied question of how to obtain interatomic potentials from liquid state structure factors $S(k)$ for simple
liquids, where the need for very accurate experimental data, and sophisticated predictor-corrector inversion
schemes~\cite{Reat86} is firmly established.  For simple liquids,  inversions are  subtle in part because the structure is dominated
by their hard-cores~\cite{Hans86}.  For example, the $S(k)$ for monotonic atomic liquids are almost quantitatively approximated by
those of hard spheres (HS)~\cite{Ashc66,Hans86} so that the attractive interactions can only be determined by measurements accurate enough to
distinguish the small deviations from HS like behaviour.

  For particles in solution, the range of  relevant effective volume fractions $\eta$ is larger than in  the  better studied case of simple liquids near
the triple point.  In particular, volume fractions are often much lower than those that lead to an entropy dominated freezing transition.  
At these lower volume fractions, and for short-range attractive interactions, the real-space structure can be dominated by the attractions, so that one can speak of {\em energetic fluids}~\cite{Loui01a}.   Extraction of effective  interactions from real-space measurements of the pair-correlation function $g(r)$ is a well-conditioned problem for energetic fluids~\cite{Roya07}.
However, real-space measurements are not always possible or  available.  In this article I will show that inversions from $S(k)$  are as
problematic for energetic fluids as they are for entropic fluids, albeit for different reasons.   These problems can be partially circumvented by studying {\em isosbestic points} -- values of the wave-vector $k$ for
which $S(k)$ is invariant under changes of the attractive potential
well depth.  Isosbestic points can be used to determine the effective range of the
$v^{eff}(r)$, and also suggest an extended corresponding states
principle for particles in solution, described by three experimentally
accessible variables: the
range defined by these points, the particle density $\rho$ and the
reduced second osmotic virial coefficient $B_2^*=B_2/B_2^{HS}$, where
$B_2^{HS}$ is the second virial coefficient of a hard-sphere (HS)
system.

%
\section{Inverting structure factors at low packing fractions}

For particles in solution, the scattering intensity $I(k)$, measured
by light, X-rays, or neutrons, is usually interpreted by dividing
$I(k)$ by the single particle form factor $P(k)$, to obtain the
structure factor $S(k) = I(k)/(\rho P(k))$~\cite{Hunt01}; here
$\rho=N/V$ is the density. From $S(k)$, inferences can then be made about the
form of the inter-particle interactions $v^{eff}(r)$.

\begin{figure}
\begin{center} \includegraphics[width=10cm]{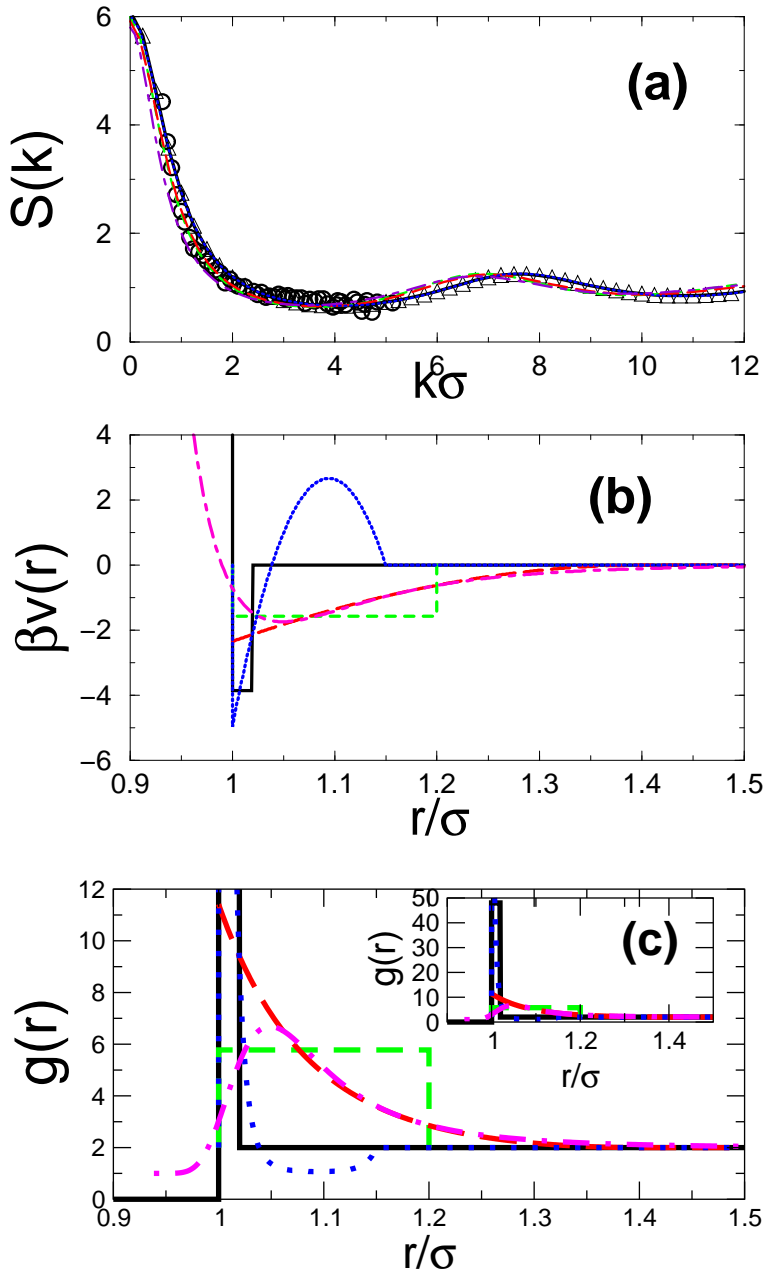}
\includegraphics[width=10cm]{Huang-combined-gofr.eps}
\end{center}
\caption{\label{fig:Huang} (colour online) A single set of experimental data can be
interpreted by many different effective potentials.
{\bf (a)} The experimental
SANS data (circles) on a microemulsion at packing fraction $\eta =
0.075$ is from \protect~\cite{Huan84}.  The open triangles denote the
$S(k)$ from the Baxter model\protect~\cite{Baxt68} for $B_2^* = -1.60$,
while the other $S(k)$ are calculated for the potentials
shown in {\bf (b)}. These include SWs with $\Delta_{SW}=0.02$ (solid
lines) and $\Delta_{SW}=0.2$ (dashed), an AO form with
$\Delta_{AO}=0.4$ (long-dashed), a generalised depletion form with
$\Delta_D=0.2$(dotted) and a LJ-12 (dot-dashed) potential.
{\bf (c)}  The radial distribution function $g(r)$ is much more sensitive to the potentials in this energetic fluid regime than the $S(k)$ is.  The line-styles in
the three graphs correspond with each other. 
 }
\end{figure}

Fig.~\ref{fig:Huang} helps set the stage and introduce the problem to
be addressed here. Experimental data~\cite{Huan84}, taken
from small-angle neutron-scattering (SANS) of a microemulsion composed
of small water droplets, coated with a layer of surfactant (AOT), is
compared to some theoretical structure factors $S(k)$.  In the
original letter~\cite{Huan84} a best fit, using an approximate integral
equation technique, yielded a hard-core diameter of $\sigma = 60 \AA$,
a packing fraction $\eta = \pi \sigma^3/6 = 0.075$, and an attractive
square well (SW) potential with a range $\Delta_{SW} = 0.02 \sigma$ and a well
depth of $\beta v(\sigma) = 3.85$.  The quality of the integral
equation fit was then confirmed by independent computer simulations.
However, as strikingly demonstrated by Fig.~\ref{fig:Huang}, a wide
range of other potentials, depicted in Fig.~\ref{fig:Huang}(b), also
lead to fits of similar accuracy.  These include HSs with
 attractive SWs of two different ranges $\Delta_{SW}$,
an Asakura-Oosawa (AO) depletion potential~\cite{Asak58} of range
$\Delta_{AO}$,  an alternate simplified depletion potential of
range $\Delta_D$ showing the effects of solvation layers~\cite{Gotz98},
as well as a Lennard-Jones-n (LJ-n) potential, defined as:
\begin{equation}\label{eq2}
 v(r) = 4 \epsilon \left(\left(\frac{\sigma}{r}\right)^{2n} -
\left(\frac{\sigma}{r}\right)^{n}\right),
\end{equation}
with $n=12$.  The $S(k)$ are calculated within the Percus Yevick (PY)
integral equation closure~\cite{Hans86,PY}, and are close to the
analytical PY solution of Baxter's model~\cite{Baxt68}.

Improving significantly on these pioneering experiments is hard.
$S(k)$ is obtained by dividing $I(k)$ through by $P(k)$, which is only
approximately determined, and which rapidly decays to zero for
increasing $k\sigma$, so that achieving better accuracies for a larger
$k\sigma$ range is very difficult. 
  As is clear from Fig.~\ref{fig:Huang} (c), real-space measurements of $g(r)$ would  be much more sensitive to the effective potentials.  
  Although increasingly accurate techniques being
developed for the real-space measurement of the $v^{eff}(r)$ of
particles in solution~\cite{Isra92,Roya07}, these are often limited to certain
particle types or sizes.   The microemulsion measured in~\cite{Huan84}, for example,  would be hard to characterise by current real-space techniques

The upshot of Fig.~\ref{fig:Huang} is that deducing an effective
potential from experimental scattering data at one state point is
difficult;  the inversion is ill-conditioned. Clearly more information
is needed to interpret the data~\cite{critique}. 

But first, what  can be inferred from a single $S(k)$?  One hint comes the
Baxter model, which is completely determined by the packing fraction
$\eta$ and the reduced osmotic virial coefficient $B_2^*$ or equivalently, the
Baxter parameter, defined as $\tau = \frac14/(1- B_2^*)$.  The $S(k)$
in Fig.~\ref{fig:Huang}, with effective $\tau$ parameters ranging from $\tau =
0.093$ to $\tau = 0.10$   are all close to the Baxter model result with
$\tau = 0.096$.   
  To first order therefore, measuring a single $S(k)$ in
this regime of low packing fraction, commonly encountered for
particles in solution, does not result in much more information about
the effective potential than the reduced second virial coefficient
$B_2^*$. Just as found for simple liquids, where inversions are hard because the
$S(k)$ resemble those of HS fluids~\cite{Hans86}, here difficulties arise because they
are close to  the Baxter $S(k)$.  Distinguishing between different potentials will only be possible by measuring  deviations from the Baxter model, and these may be very small at any one statepoint.

\section{Isosbestic points in structure factors}

\begin{figure}
\begin{center} \includegraphics[width=10cm]{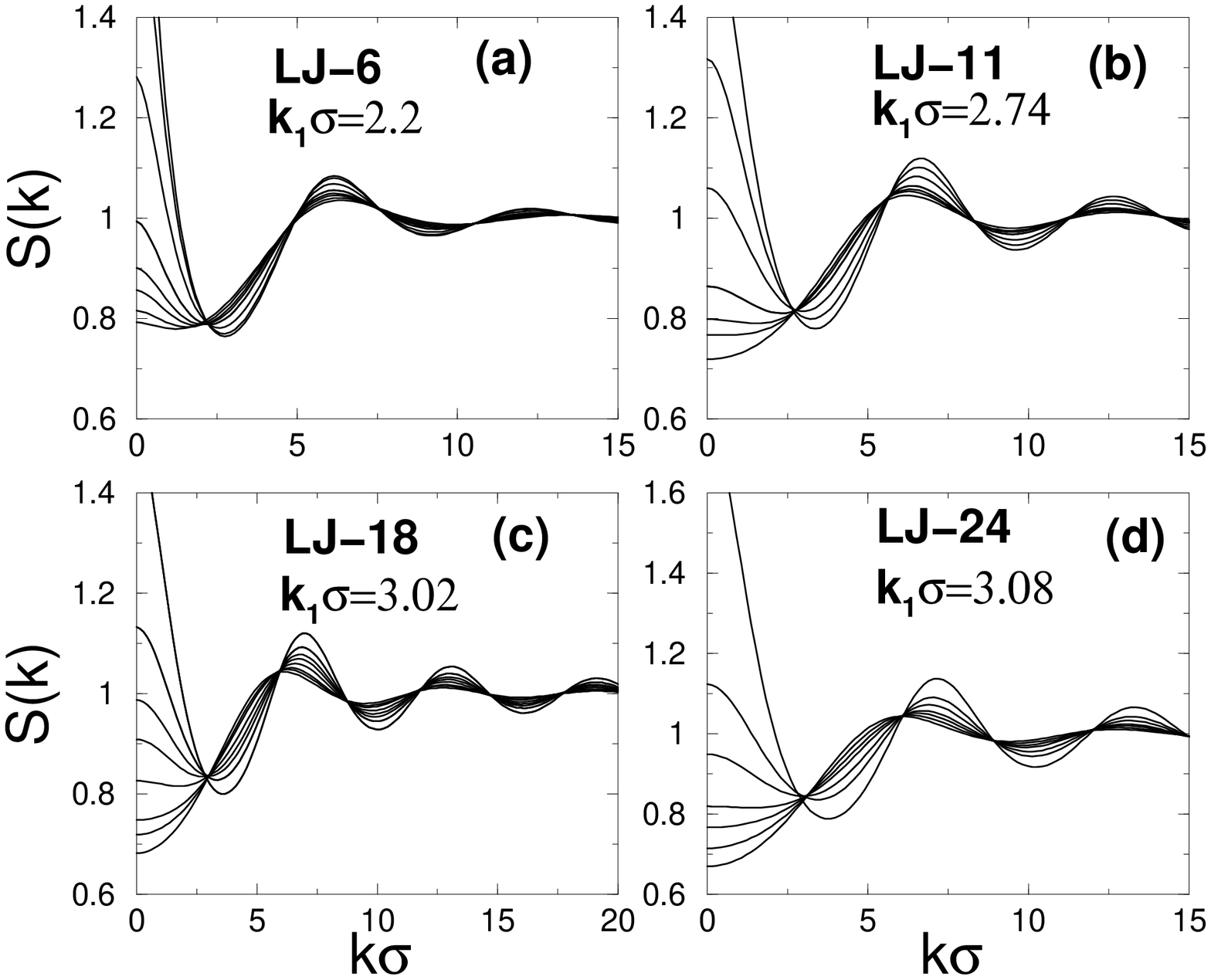}
\includegraphics[width=10cm]{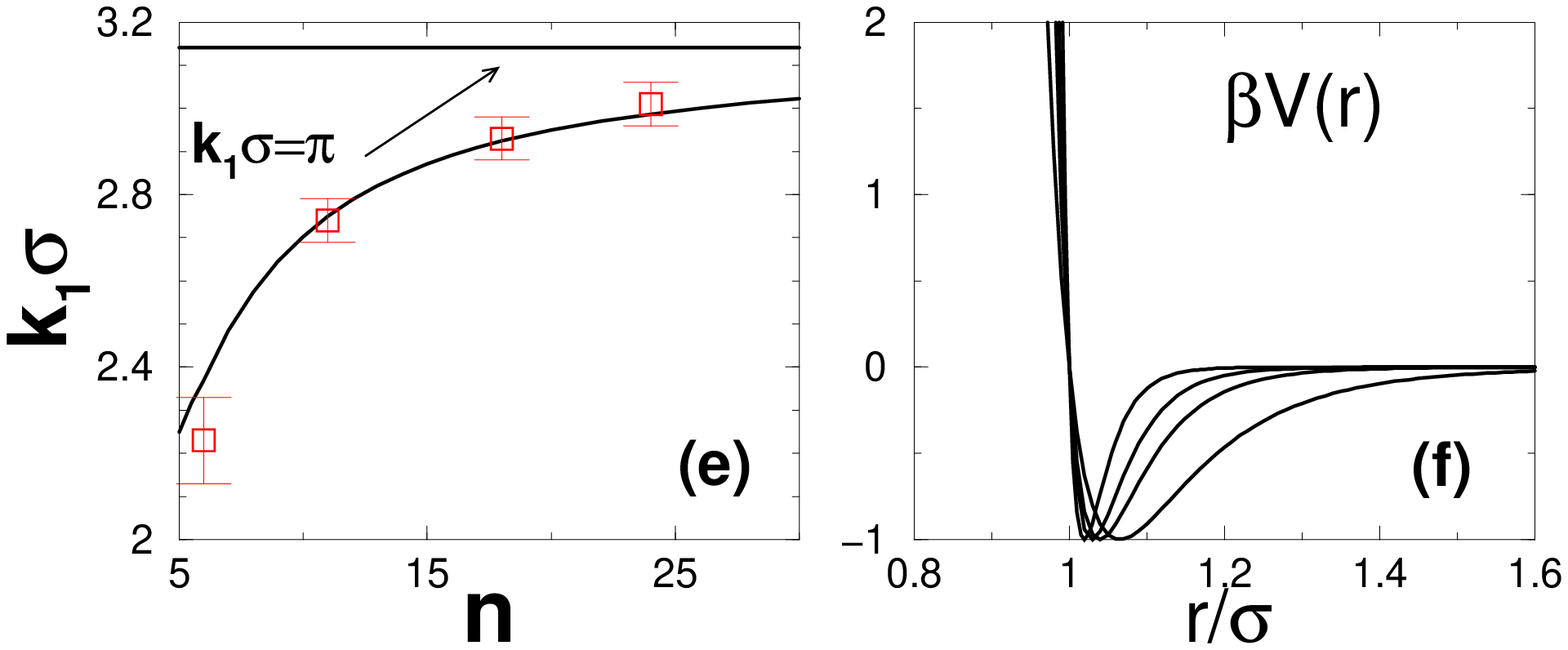}
\end{center} 
\caption{\label{fig:Isosbestic} (colour online) (a)-(d) $S(k)$ were calculated at $\eta
= 0.0576$ and at different temperatures for each of the four LJ-n
potentials $v^{eff}(r)$ given by Eq.~(\protect\ref{eq2}).  Different
$v^{eff}(r)$ lead to different isosbestic points $k_n\sigma$.  (e) The
first isosbestic point $k_1\sigma$ approaches $\pi$ with increasing
$n$. A simple theory (solid line), described in the text, fits the
data well. (f) The range of the  LJ-n potentials  in (a)-(d) 
decreases with increasing $n$.\vspace*{-0.0cm}}
\end{figure}

To make further progress, measurements at other state points are
necessary.  This is done in Figs.~\ref{fig:Isosbestic}(a)-(d), which
depict the $S(k)$ calculated for four LJ-n potentials at different
temperatures.  A best fit to the Baxter model $S(k)$ at each
temperature would result in $B_2^*$ as a function of temperature.  But
there is clearly more information in these curves: Each of the four
$v^{eff}(r)$ results in a different set of {\em isosbestic points},
where the $S(k)$ is invariant for different temperatures.
Fig.~\ref{fig:Isosbestic}(e) shows that the first isosbestic point
$k_1\sigma$ increases with increasing $n$, reflecting the decrease of
the range of the LJ-n potential~(\ref{eq2}) depicted in
Fig.~\ref{fig:Isosbestic}(f). (Error bars in
Fig~\ref{fig:Isosbestic}(e) reflect the approximate nature of 
the isosbestic points.)

These observations can be quite easily rationalised by the following
simple theory: If the total correlation function $h(r)=g(r)-1$ is
split into two parts, with $h_0(r) = -1$ for $r < \sigma$, and $h_1(r)
= g(r)-1$ for $r \geq \sigma$, then the structure factor $S(k) = 1 + \rho \hat{h}(k)$ simplifies to
\begin{equation}\label{eq3}
S(k) = 1 + \frac {4 \pi \rho}{k} j_1(k) + \rho \hat{h}_1(k),
\end{equation}
where $j_1(k)$ is the first spherical Bessel function, and
$\hat{h}_1(k)$ is the Fourier Transform (FT) of $h_1(r)$.  For
potentials with a hard-core, such as most of those depicted in
Fig.~\ref{fig:Huang}(b), this is exact, but even for the LJ-n
potentials this is a good approximation~\cite{sigma}.  Since
$\hat{h}_0(k) = \frac{4 \pi \rho}{k} j_1(k)$ is
independent of temperature (barring small effective $\sigma$ effects),
Eq.~\ref{eq3} suggests that isosbestic points occur whenever
$\hat{h}_1(k)=0$.  We note that a similar derivation was made within the MSA approximation by Tuinier and Vliegenthart~\cite{Tuin02}, who showed that isosbestic points may also occur for longer ranged potentials.

\begin{figure}
\begin{center} \includegraphics[width=10cm]{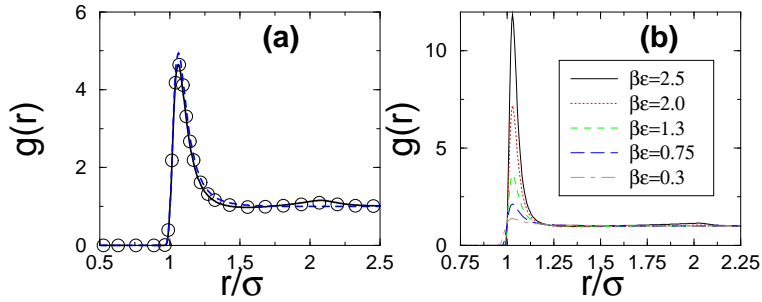}
\end{center} 
\caption{\label{fig:gofr} (colour online) (a) The PY approximation (solid line) very
accurately reproduces these molecular dynamics simulations~\cite{PY} of
a LJ-12 potential at $\beta \epsilon = 1.6$, and $\eta=0.1$.  The
simpler $g(r) = \exp (-\beta v(r))$ form (dashed lines) is also
accurate. (b) When the temperature is changed, the dominant effect on
$g(r)$, calculated here with PY, is to change its amplitude.  }
\vspace*{-0.0cm}\end{figure}

It has already been pointed out that in many experimentally relevant
cases with low $\eta$, the $g(r)$ are surprisingly well approximated
by a simple form $g(r) = \exp (- \beta v(r))$~\cite{Loui01a},
as explicitly demonstrated in Fig.~\ref{fig:gofr}. This
simple approximation, equivalent to taking a Mayer
function~\cite{Hans86} for $h(r)$, works best for particles with short
range attractive potentials. These can become quite deep, leading to
large values of $g(r)$ near contact, well before the system crosses a
liquid-liquid or liquid-solid phase-line.  Within this approximation,
the dominant effect of varying the temperature is to change the
amplitude of $h_1(r)$ (as demonstrated in Fig.~\ref{fig:gofr}(b)). To
first order, the period of $\hat{h}_1(k)$ is not affected and each
$k\sigma$ where $\hat{h}_1(k)=0$ leads to an isosbestic point. For an
infinitely narrow potential, the isosbestic points would be at $k_n
\sigma = n\pi$, but for a finite potential range there is a
phase-shift, which explains why the $k_n\sigma$ move progressively
further from $n \pi$ with increasing range.
  The excellent accuracy of this simple Mayer function theory for the first
isosbestic point $k_1\sigma$ is
demonstrated by the solid line in Fig.~\ref{fig:Isosbestic}(e).

\begin{figure}
\begin{center} \includegraphics[width=10cm]{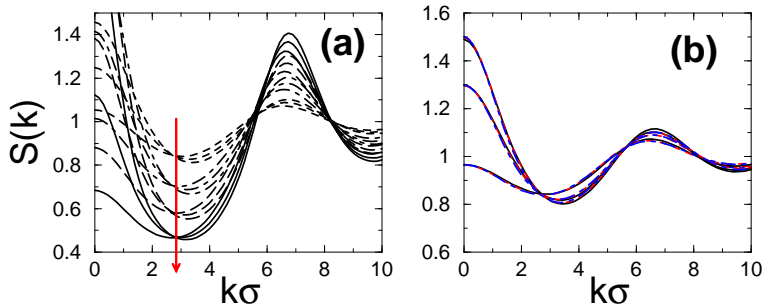}
\end{center} 
\caption{\label{fig:sofk}(colour online) Fig (a) The position of isosbestic points
barely varies with density for LJ-12, shown here for three
temperatures $\beta \epsilon = 1.5,1.33,1.09$ at each density
$\eta=0.05,0.1,0.15,0.2$. The line with the arrow denotes $k_1\sigma$
for increasing $\eta$.  (b) Three potentials leading to the same isosbestic
points: SW ($\Delta_{SW} = 0.26$) (solid lines), AO ($\Delta_{AO}=0.6$)
(dashed lines) and LJ-22 (dot-dashed lines) for $\eta=0.05$ and
$B_2^*=0, -0.719, -1$ (plots in descending  order of $B_2$ at $k=0$). }
\vspace*{-0.0cm}\end{figure}

The examples depicted in Fig.~\ref{fig:Isosbestic} are at a relatively
low packing fraction, but the isosbestic points are robust up to
packing fractions of at least $\eta = 0.2$, as shown in
Fig.~\ref{fig:sofk}(a)~\cite{density}. These points are accurately
described by the simple theory, described above.

Effective potentials in complex fluids may have many origins~\cite{Liko01}, so that one cannot simply vary the temperature to change the well depth.  Nevertheless, the analysis above suggests that 
if varying a parameter leads to isosbestic points, then it is possible that 
that the well-depth of $v^{eff}(r)$ is changing while the range is
not. When combined with the variation of $B_2$, this can help fix the
form of $v^{eff}(r)$.  However, this information may still not be
sufficient, since the well-depth of $v^{eff}(r)$ may depend in an
(unknown) non-linear fashion on the parameters like the temperature
$T$, the pH, and salt or other additive concentration.  Furthermore, as
shown in Fig.~\ref{fig:sofk}(b), different potential shapes, picked to
have the same isosbestic points, can generate similar $S(k)$ for each
value of $B_2^*$.

\section{An extended corresponding states principle}

This similarity of the $S(k)$ ties in with the work of Noro and
Frenkel (NF), who proposed that many properties of particles in
solution could be understood from an extended corresponding state
principle based on the variables $\eta$, $B_2^*$, and an effective well
depth $\beta \epsilon$~\cite{Noro00}.  However, the three potentials
shown in Fig.~\ref{fig:sofk}(b) do not have the same well-depth, while
still showing similar $S(k)$.  This suggests an alternative
corresponding states principle based on the variables $\eta$, $B_2^*$,
and an effective range $\Delta_{eff}$ related to the isosbestic
points. In fact, as NF point out, the potential range is not always
uniquely defined when comparing different $v^{eff}(r)$'s. They
suggested a non-linear mapping based on $B_2^*$ to derive an effective
SW range for each potential they consider. 

\begin{figure}
\begin{center} \includegraphics[width=8cm]{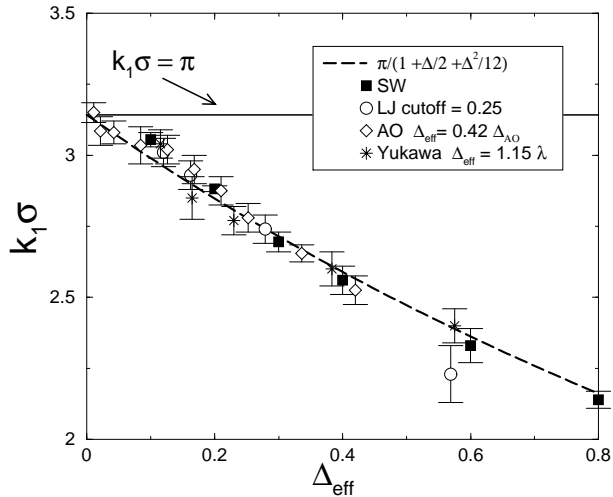}
\end{center} 
\caption{\label{fig:univ} Defining effective range parameters from
isosbestic points. For each potential type a simple mapping described in the text was used to derive 
an effective square well range $\Delta_{eff}$.   For each potential, a PY calculation was used to find the isosbestic points, and these are shown in the plot for four different kinds of potentials.
The long-dashed line denotes an accurate analytical
approximation for the isosbestic points of square-well fluids. If one does not know what the potential is, then   $\Delta_{eff}$ can be directly estimated
from the $k_1 \sigma$ measured in experiments.    }
\vspace*{-0.0cm}\end{figure}

Here I define the effective  range $\Delta_{eff}$ of a given potential as being that of an equivalent SW fluid with the same 
same $k_1 \sigma$.   A simple and 
accurate approximation of $k_1 \sigma$ for the SW fluid at low concentrations is given
by  $k_1 \sigma \approx \pi/(1 + \Delta_{eff}/2 + \Delta_{eff}^2/12)$
which follows from an expansion of the exact solution to the simple
theory of isosbestic points.  Thus the value of the effective range is then well approximated by:
\begin{equation}
\Delta_{eff} = 24 \left(\frac{\pi}{3 k_1 \sigma} - \frac{1}{12}\right)^\frac12 - 12.
\end{equation}
By comparing the $k_1 \sigma$ values for different potentials, some simple empirical rules can be derived:
For the LJ-n
potentials this criterion reduces to the distance $r-\sigma$ where
$\beta v(r)$ reaches $1/4$ of its maximum depth, for the AO potential
the effective SW range is $\Delta_{eff} \simeq 0.42 \Delta_{AO}$, and
for the hard-core Yukawa potential, with an attractive tail of the
form $ v(r) = \epsilon \exp(-r/\lambda)$ for $r>\sigma$, the mapping
is $\Delta_{eff} \simeq 1.15 \lambda$.  The results are depicted in
Fig.~\ref{fig:univ}, and show that these simple linear mappings
successfully define an effective range for each potential type.   

Another argument for including the $\Delta_{eff}$ (instead of the
well-depth) as one of the relevant $3$ variables is that the range
determines the topology of the phase-diagram~\cite{Noro00}.  The
critical range below which the fluid-fluid transition becomes
metastable to the fluid-solid line is at $\Delta_{eff} \approx 0.15$
for all four potentials (SW, AO, Yukawa and LJ-n), something
consistent with what NF found.  In addition, $\Delta_{eff}$ is
directly experimentally accessible through $k_1\sigma$.  For example,
taking advantage of the proximity of a metastable fluid-fluid critical
point to enhance nucleation, a proposal made for protein
solutions~\cite{tenW97}, would involve tailoring solution conditions
such that the measured isosbestic point is just above $k_1\sigma =
2.92$, so that $\Delta_{eff}$ is just below $0.15$.

\section{Discussion}

In summary then, extracting effective potentials $v^{eff}(r)$ from
$S(k)$ at the lower packing fractions typically encountered for
particles in solution is a poorly conditioned problem. Whereas in
simple liquids the HS model describes the dominant features of the
$S(k)$, here the Baxter model appears to be the fundamental underlying
model for $S(k)$ around which different attractive  potentials  only induce
a mild perturbation. More information can be obtained by studying the
isosbestic points, which  help determine the range $\Delta_{eff}$ of the
potentials. Together with $\eta$ and $B_2^*$, $\Delta_{eff}$ can be
used to define an extended corresponding states principle.  Particles
in solution with similar values of these three parameters should have
similar properties, such as the relative stability of the fluid-fluid
and fluid-solid binodals.  In other words, many solution properties
could be deduced without having to actually invert to an explicit form
of $v^{eff}(r)$.  

Nevertheless, It should always be kept in mind that the effective pair potential that reproduces the structure, even though it is unique~\cite{Hend74}, may not be the same as the effective potential that reproduces certain thermodynamic properties.  Such representability problems~\cite{Loui02b} are typically important when there are large three-body interactions, or when the underlying system has anisotropic interactions~\cite{John07}.   For systems with truly short-ranged potentials, one doesn't expect strong many-body effects.  Nevertheless,  anisotropic interactions may be common for colloidal suspensions. They lead to important deviations from spherically symmetric potentials~\cite{Alla03,Noya07,Roma10,Bian11}, could lead to representability problems if a potential is inferred from inverting structure factors.  

 In addition,  particles in solution are often polydisperse, an effect that has been ignored in the current analysis.  Polydispersity can manifest in multiple properties, i.e. in the effective diameters as well as in the interaction strengths. Clearly this adds a different dimension of complexity, and it would be interesting to see how isosbestic points would be affected by  polydispersity.  Encouragingly, structure factors have been solved for polydispersity in systems like the Baxter model~\cite{Gazz03}, so that progress may be expected on this front in the future.

Finally,  numerous examples of
isosbestic points in $S(k)$  can be found in the literature, ranging from
colloids with short-range sticky coats~\cite{Duit91}, to
colloid-polymer mixtures~\cite{Ye96}, to globular
proteins~\cite{Tard99}, to magnetic colloids~\cite{Dubo99} and to
colloid-micelle mixtures~\cite{Bart}. Interestingly, these points
should also appear in $I(k)$.

\section*{Acknowledgements}

The author is  very pleased to contribute to this Special Issue of Molecular Physics dedicated to Professor Luciano Reatto in recognition of his important contributions to the field, and wishes him much success in the coming years.  The author also thanks P. Bartlett, J. P. K. Doye for valuable discussions.     He also thanks Remco Tuinier and Gerrit A. Vliegenthart for sharing with him their independent discovery of isosbestic points in structure factors~\cite{Tuin02}.


\begin{thebibliography}{99}
\bibitem{critique} I should add that this is not an implicit critique of
ref~\cite{Huan84}, where other physical considerations were indeed used
to infer the suggested short ranged potential.
\bibitem{sigma} There are several similar ways to define an effective
$\sigma_{eff}$ for a LJ-n potential, see e.g.\protect~\cite{Hans86}.
A more careful analysis shows that for our examples, these don't differ much from
$\sigma$, and  have a marginal effect on the isosbestic points.
\bibitem{density} The isosbestic points typically
move to a marginally higher $k$ with increasing $\eta$, an effect also
seen for the Baxter model;  this can be used for a small
correction to isosbestic points.
\end{thebibliography}
\end{document}